\documentclass{article}
\usepackage{spconf,amsmath,graphicx}

\usepackage{array}
\usepackage{xcolor}
\usepackage{url}
\newcolumntype{P}[1]{>{\centering\arraybackslash}p{#1}}

\title{Extracting speaker and emotion information from self-supervised speech models via channel-wise correlations}
\name{\resizebox{\linewidth}{!}{\vspace{-1cm} Themos Stafylakis$^{1\dag}$, Ladislav Mo\v{s}ner$^{2\dag}$, Sofoklis Kakouros$^{3\dag}$, Old\v{r}ich Plchot$^{2}$, Luká\v{s} Burget$^{2}$, Jan \v{C}ernocký$^{2}$}}
\address{
$^1$Omilia - Conversational Intelligence, Athens, Greece\\
$^2$Brno University of Technology, Faculty of Information Technology, Speech@FIT, Czechia \\
$^3$University of Helsinki, Finland \\
}

\copyrightnotice{978-1-6654-7189-3/22/\$31.00~\copyright2023 IEEE}
\begin{document}
%
\maketitle
\begin{abstract}
Self-supervised learning of speech representations from large amounts of unlabeled data has enabled state-of-the-art results in several speech processing tasks. Aggregating these speech representations across time is typically approached by using descriptive statistics, and in particular, using the first- and second-order statistics of representation coefficients. In this paper, we examine an alternative way of extracting speaker and emotion information from self-supervised trained models, based on the correlations between the coefficients of the representations --- correlation pooling. We show improvements over mean pooling and further gains when the pooling methods are combined via fusion. The code is available at \url{github.com/Lamomal/s3prl_correlation}.
\let\thefootnote\relax\footnotetext{\dag Equal contribution} \end{abstract}
\section{Introduction}
\label{sec:intro}

Large speech models trained in a self-supervised manner, such as Wav2Vec 2.0, HuBERT, and WavLM, have shown exceptional performance when finetuned on the downstream tasks \cite{hsu2021hubert,fan2020exploring,chen2021wavlm}. However, finetuning the weights of such models on each task is a non-scalable solution for production systems performing several of these downstream tasks in real-time. For such systems, the preferable solution would be to extract a single set of speech features from a shared model, followed by a task-specific lightweight classifier.

To this end, the Speech processing Universal PERformance Benchmark (SUPERB) challenge \cite{yang2021superb} was recently introduced, with the goal of benchmarking the performance of such speech models on a variety of speech tasks (e.g. ASR, keyword spotting, spoken language understanding, emotion recognition, speaker recognition, identification and diarization, a.o.) keeping the models' weights frozen and employing a lightweight task-specific classifier \cite{yang2021superb}.    

Among the interesting findings of SUPERB is the ability of such models to encode speaker and emotion information in the intermediate layers. The models are trained using an implicitly phonetic loss (typically a masked-language model style loss over quantized vector representations). It means that speaker and emotion modeling is not directly encouraged. However, the models must be performing some kind of speaker and emotion modeling in the intermediate layers, in order to suppress this nuisance variability in the output layer.

For those tasks requiring {\it utterance-level} classification or representation learning, the SUPERB benchmark employs a lightweight trainable classifier incorporating pooling. The pooling methods used are (a) mean pooling, and (b) statistics pooling (concatenated mean and standard deviation, \emph{std}, vectors) in speaker verification (SV), which is the standard pooling method for x-vectors~\cite{snyder2018xvec}. 

Although mean pooling typically yields good performance in several speaker and emotion modeling tasks \cite{wang2021pooling}, we should emphasize a crucial difference between these models and the SUPERB setup \cite{yang2021superb}; the fact that the models are frozen, after being pretrained using a loss function that does not directly encourage speaker or emotion modeling. As a result,  methods using mean pooling make simplifying assumptions about the way the information is encoded into the network's internal representations. 

Mean pooling, including the statistics pooling variant, assumes that the correlations between different feature dimensions (or \emph{channels}) are of little or no importance. This might be true if the model is trained or finetuned this way, i.e. with a classifier that extracts fixed-length representations via mean pooling (e.g. using a softmax classifier with cross-entropy loss). After all, deep neural networks have the capacity to extract information relevant to the task in a useful form. However, in the case of pretrained models, such an assumption is at least questionable, and methods that take into account correlations between feature channels should be considered. 

In this paper, we focus on three SUPERB tasks that require sentence-level representations, namely speaker verification and identification, and emotion recognition. We show that a significant portion of the information related to speaker and emotion is encoded in the channel-wise correlations of the intermediate layers. The idea is based on the \emph{correlation pooling} method originally introduced in \cite{stafylakis2021speaker}. However, the network in \cite{stafylakis2021speaker}, apart from having a very different architecture (2D-ResNet), was not pretrained and/or frozen, but trained from scratch for the speaker recognition task.  

\section{RELATED WORK}
\label{sec:related_work}

Correlations between channels have been explored in computer vision, as a means to extract and/or modify the style and the texture of images. The work of L.A. Gatys {\it et al.} \cite{gatys2016image} introduced neural-style transfer, showing that such image characteristics are captured by channel-wise correlations from Deep Convolutional Nets (ConvNets) trained for object recognition using ImageNet. 
Their method was adapted to speech generation and voice conversion system in \cite{chorowski2018using}, where the authors demonstrated that intermediate layer representations encode speaker characteristics.

The proposed pooling method was first introduced in \cite{stafylakis2021speaker}. It was shown that 2D Deep Convolutional Nets (ResNet-34) can be trained from scratch while using channel-wise correlation pooling for frequency ranges, and the experiments on VoxCeleb demonstrated improvements over the standard statistics (mean-std) pooling. In this work, we extend this method to pretrained self-supervised transformer models (which are 1D since self-attention operates only across the temporal axis) and we also test it on an emotion recognition. 

\section{Sentence-level representations in SUPERB}
\label{sec:SUPERB}
In this section, we describe the proposed correlation pooling as well as several details related to the SUPERB challenge. 

\subsection{Transformer-based architectures}
The most powerful self-supervised models follow the transformer architecture. The input features are extracted from the waveform (at a rate of 50\,fps) via a ConvNet, which is trained jointly with the transformer. The ConvNet is typically frozen even in cases where the model is finetuned, as it can easily overfit. A transformer layer follows the encoder block architecture defined in \cite{vaswani2017transformer}. It consists of a multi-head self-attention layer, followed by a feed-forward layer, while layer-normalization is added in both layers. Critically, skip connections are added between these layers, as in ResNets. An interesting property of architectures equipped with skip connections is that the correspondence between units of the representations of different layers is preserved (i.e. the representations are aligned). Each layer adds further contextualization (via self-attention in the case of transformers) and details needed for optimizing the task defined by the loss function (e.g. modeling and subsequently suppressing nuisance variabilities, such as speaker, noise, channel, and emotion). However, the $i$th unit of the $l$th layer's representation captures a similar characteristic with the $i$th unit of all other layers.

\subsection{Layerwise pooling}
The alignment between representations from different layers permits an easy way of extracting information relevant to the downstream task from all (i.e. both output and intermediate) representations, by collapsing them into a single one via a weighted average. The weights are learned jointly with the task-specific classification network. More concretely, the averaged representation for a transformer with $L$ layers is expressed as

\begin{equation}
\label{eq:layerwise}
{\bf h}_t = \sum_{l=0}^{L}\gamma_{l}{\bf h}_{t,l}, 
\end{equation}
where the weights $\sum_{l=0}^{L}\gamma_{l}=1, \, \gamma_{l} \geq 0$ 
are implemented with a learnable vector of size $L+1$, followed by the softmax function, and ${\bf h}_{t,l}$ is the representation of the $l$th layer at time $t$ (${\bf h}_{t,0}$ is the output of the ConvNet). 

Note that collapsing all $L+1$ representations into a single one via a simple weighted averaging would not make sense for networks without skip connections, even if their sizes were the same, unless the models were trained with a loss defined on layerwise-averaged representation (which is not the case here, since the loss of the self-supervised models is applied to the output layer). Exploring all representations would require either concatenation along the feature dimension (increasing the size of the latter by a factor or $L+1$) or an exhaustive search (i.e. training a different classifier for each of the $L+1$ layers) for finding the single most informative representation for each task. 

The SUPERB protocol suggests the weighted-average type of layer-wise pooling for evaluating different models and tasks, and so do we in this work. Note that this type of layer-wise pooling was also used in ELMo, a bidirectional LSTM-based language model with skip connections, designed to extract deep contextualized word representations~\cite{peters-etal-2018-deep}. 
\subsection{Mean pooling}
\label{ss:aver}
Tasks requiring a sentence-level classification typically employ a pooling method, such as mean, max or attentive pooling. Mean pooling, which is employed in SUPERB is defined as
\begin{equation}
    {\bf r} = \bar{\bf h} = \frac{1}{T}\sum_{t=1}^{T}{\bf h}_{t}, 
\end{equation}
where $T$ is the number of acoustic features of an utterance extracted by the ConvNet, ${\bf r}$ is the resulting pooled representation, while ${\bf h}_{t}$ are the representations at time $t$ after layer-wise pooling. Concatenating the pooled representations with std features (statistics pooling) is in general helpful in speaker recognition \cite{wang2021pooling}, and is implemented as
\begin{equation}
    {\bf r} = \left[\bar{\bf h} ; \left(\frac{1}{T}\sum_{t=1}^{T}({\bf h}_{t} - \bar{\bf h})^2 \right)^{1/2}\right], 
\end{equation}
where $[\cdot;\cdot]$ denotes vector concatenation and the exponents should be considered as element-wise operators.

The representation ${\bf r}$ is optionally projected onto a lower-dimensional space via a learnable linear layer resulting in an internal representation (or embedding) ${\bf w}$. What follows is a classification head, composed of a linear layer with an output size equal to the number of classes (speaker identities, emotion types), and the softmax function.

\section{Correlation pooling}
\label{sec:correlation}
\subsection{Motivation}
A drawback of mean pooling is that it ignores correlations between different dimensions (channels) in ${\bf h}$. Research in computer vision has demonstrated that certain characteristics of an image, such as style and texture, are better captured by these correlations, especially when the model is trained to predict other characteristics such as objects (e.g. ImageNet) \cite{gatys2016image}. By analogy with computer vision, we speculate that utterance-level characteristics, such as speaker, emotion (but also channel, noise, a.o. not considered in SUPERB) are to a large extent encoded in channel-wise correlations in speech models, which are trained with an objective that encourages suppressing them in the output layer. Note that in HuBERT and WavLM, the discrete acoustic units are highly correlated with phonemes (as shown in \cite{hsu2021hubert}) while the output layer carries little or negligible speaker information in Wav2Vec 2.0 and HuBERT \cite{novoselov2022robust,van2022comparison}.

\subsection{Implementation}
We re-implement the pooling method introduced in \cite{stafylakis2021speaker}, by adapting it to 2D tensor representations (i.e. channel and time) as opposed to 3D (i.e. channel, frequency, and time). We first project ${\bf h}$ onto a lower-dimensional space via a linear projection, and then apply mean and variance normalization over time (i.e. standarization) resulting in ${\bf o}$. We then calculate the outer products and apply mean pooling, resulting in the correlation matrix of ${\bf o}$, i.e.
\begin{equation}
    {\bf C} = \frac{1}{T}\sum_{t=1}^{T}{\bf o}_{t}{\bf o}'_{t}, 
\end{equation}
where ${\bf o}'_{t}$ is the transpose of ${\bf o}_{t}$ (note that ${\bf o}_{t}$ is a column vector). Since the matrix is symmetric and its diagonal elements are equal to 1, we vectorize
the elements above the diagonal and optionally project them onto a linear layer, resulting in the embedding ${\bf w}$. 

As a regularization method, we use \emph{channel dropout}, i.e. we zero out whole channels (with probability equal to 0.25) before estimating the correlation matrix. It results in zero rows and columns in ${\bf C}$. This method appears to be effective in SV, where the goal is generalization to unseen speakers~\cite{stafylakis2021speaker}.   

\section{EXPERIMENTS}
\label{sec:experiments}
We conducted experiments in speaker identification (SID), speaker verification (SV), and emotion recognition (ER) using the SUPERB protocol, while all experiments and implementations are based on the \emph{s3prl toolkit}\footnote{\url{https://github.com/s3prl/s3prl}}. It is worth noting that contrary to other SUPERB systems, \emph{no data augmentation} was applied during training. Tasks, models, and results are described in detail in the next subsections.

\subsection{Speaker identification (SID)}
\label{ssec:speakerid}

The first utterance-level task we examine is speaker identification (SID). As opposed to speaker verification (SV), SID focuses on closed-set speaker classification. 

\subsubsection{Architectural details}
\label{ssec:sidimplementation}
We note that we follow the architecture of the SID model defined by the s3prl toolkit. Per-layer representations are aggregated by a weighted average, where the weights are optimized along with other parameters during training. Obtained frame-level features are projected to a 256-dimensional space. What follows is pooling over time. The SUPERB benchmark suggests employing mean pooling. In contrast, motivated by a common practice in SV, we also experiment with statistics pooling. As a last method, we employ correlation pooling. The resulting utterance-level representation is an input to a classification head. A standard cross-entropy (CE) is optimized during training.

To regularize correlation pooling we employ dropout, which we apply before normalization and the pooling itself with a probability of 0.25. As opposed to the standard dropout, entire channels are dropped (also used in \cite{stafylakis2021speaker}).

\subsubsection{SID Dataset}
\label{ssec:siddataset}
For both training and evaluation of SID systems, we utilize the VoxCeleb 1 dataset \cite{nagrani2017voxceleb}. It comprises 1,251 identities. The number of utterances totals 153,516. We follow exactly the partitioning to training, development, and test sets defined by the authors\footnote{\url{www.robots.ox.ac.uk/~vgg/data/voxceleb/vox1.html}} which is also adopted by the s3prl toolkit. Out of all VoxCeleb 1 recordings, 90.1\% are allocated for training, 4.5\% for development, and 5.4\% for testing.

\subsubsection{SID Results}
The results are given in Table \ref{tab:sidsvresults} (SID column). When HuBERT is used as a backbone network, correlation pooling yields about 5\% absolute improvement compared to mean-std pooling and about 6\% compared to mean-only pooling (i.e. the one used in the SUPERB challenge). A similar boost in performance is observed when WavLM is used, where correlation pooling attains 97.7\% accuracy, as opposed to 93.0\% of mean and 94.9\% mean-std pooling. In other words, correlation pooling yields an approximate 50\% relative improvement in classification error. Finally, further improvements can be attained when correlation pooling is fused with mean and/or mean-std pooling (by averaging the logits of the two models). In Fig. \ref{fig:sidhubertwavlmweights}, we demonstrate learned layer weights for all the pooling methods.

In addition to the superior evaluation accuracy of models utilizing correlation pooling, the training converges much faster (requires about 60\% of the epochs needed for mean pooling), making the proposed method attractive also w.r.t. the computational cost.

\begin{flushleft}
\begin{table}[]
\caption{Results in SID and SV following the training and evaluation protocol defined by SUPERB.}
\label{tab:sidsvresults}
\centering
\begin{tabular}{|p{3.3cm}||P{1.9cm}|P{1.9cm}|}
 \hline
 \multicolumn{3}{|c|}{HuBERT Large} \\
 \hline
 Pooling method & SID [acc. \%] & SV [EER \%] \\
 \hline
 mean   & 89.3 & -- \\
 meanstd & 90.6 & 6.2 \\
 correlation & 95.5 & 5.3 \\
 corr. w/ dropout & 95.3 & 4.8 \\
 mean+correlation & 96.2 & -- \\
 meanstd+correlation & 96.3 & 5.1 \\
 meanstd+corr. w/ drop. & 96.3 & 4.8 \\
 \hline
 \multicolumn{3}{|c|}{WavLM Large} \\
 \hline
 Pooling method& SID & SV \\
 \hline
 mean   & 93.0 & -- \\
 meanstd & 94.9 & 4.8 \\
 correlation & 97.7 & 4.5 \\
 corr. w/ dropout & 97.7 & 4.1 \\
 mean+correlation & 97.7  & -- \\
 meanstd+correlation & 98.2 & 4.0 \\
 meanstd+corr. w/ drop. & 98.6 & 3.8 \\
\hline
\end{tabular}%
\end{table}
\end{flushleft}

\subsection{Speaker verification (SV)}
\label{ssec:speakerver}

Given two recordings (or sets of recordings), the goal of speaker verification (SV) is to determine whether the identities of the corresponding speakers match or not. As demonstrated by a long history of the evaluation series organized by NIST \cite{sadjadi2021nist, greenberg2020nistevals}, the SV is a significant utterance-level speech processing task with numerous applications, such as call center authentication, audio retrieval, forensics, and others. 

State-of-the-art systems employ embedding extractors providing summary utterance-wise speaker-descriptive vectors given features (such as log Mel-filter bank energies) extracted from the audio signal. The field of SV has been increasingly embracing pre-trained models to provide frame-level features \cite{lavrentyeva2022stc,vaessen2022fine}. It has been shown that fine-tuning the pre-trained model on Voxceleb 2, along with the embedding extractor with a speaker-discriminative objective leads to unprecedented performance \cite{chen2021wavlm} on standard benchmarks such as the Voxceleb 1 test sets. \cite{chung2018voxceleb2}.

It is worth stressing that we follow the SUPERB benchmark protocol in our study. Therefore, we keep the parameters of the pre-trained model frozen, we use only the development set of VoxCeleb 1 for training, and we train a TDNN using layer-wise aggregated features from the pre-trained model. Here, we merely focus on the effectiveness of the proposed correlation pooling w.r.t. the standard mean-std pooling used in SUPERB.

\subsubsection{Architectural details}
\label{ssec:svimplementation}
As well as in the case of SID, the architecture follows the SUPERB design. Per-layer representations are aggregated in the same way as in SID. They are subsequently projected to a 512-dimensional space. The frame-level features serve as an input to the TDNN-based (x-vector-like) model \cite{snyder2018xvec}. As in the original architecture (in which statistics pooling is employed), the last frame-wise layer produces 1500-dimensional outputs. To overcome the unbearable increase of parameters, we decrease the number of dimensions to 512 when using the correlation pooling. The 512-dimensional outputs of the first layer after pooling are used as speaker embeddings ${\bf w}$. The downstream network is trained with the additive margin (AM) softmax objective \cite{wang2018am}, where a scale of 30 and a margin of 0.4 are used (i.e. the default loss and hyperparameters of SUPERB without any optimization or data augmentation).

\subsubsection{SV Dataset}
\label{ssec:svdataset}
As in SID, we utilize the VoxCeleb 1 dataset for SV. A standard verification split, adopted also by s3prl, is followed. An original (i.e. not cleaned) version of a trial list is used for evaluation. 

\subsubsection{SV Results}
\begin{figure}[t]
  \centering
  \vspace{-5mm}
  \includegraphics[width=\linewidth]{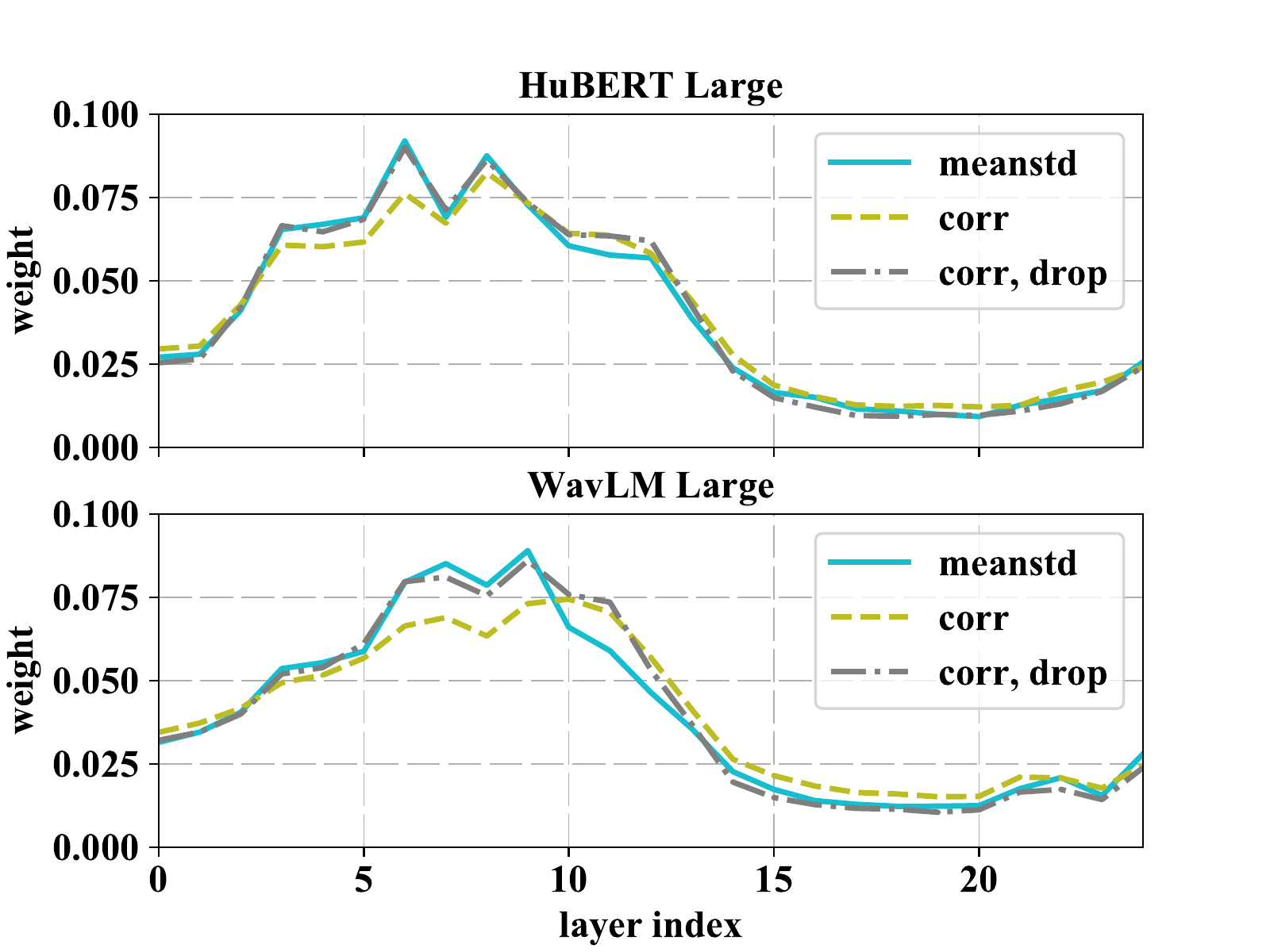}
  \vspace{-8mm}
  \caption{Learned weights per HuBERT-Large and WavLM-Large layers for SV.}
  \label{fig:svhubertwavlmweights}
  \vspace{-5mm}
\end{figure}

The results are given in Table \ref{tab:sidsvresults} (SV column). With HuBERT, correlation pooling yields a significant reduction in EER, from 6.2\% (attained with mean-std) to 5.3\% and 4.8\% without and with channel dropout, respectively. When WavLM is used, a 0.5\% absolute improvement is attained. Finally, a score-level fusion of the two pooling methods yields further improvements. We should mention that these experiments underline the importance of channel dropout as a means to prevent overfitting to the training speakers. 
We also depict the learned weights of the layer-wise pooling stage for each (output frame-level) pooling and pretrained model in Fig. \ref{fig:svhubertwavlmweights}. As we observe, the weights are similar for all pooling methods, probably due to the use of a TDNN model in the back-end.

The faster convergence of models employing correlation pooling observed in the SID experiments also holds for speaker embedding extractors. Empirically, only two-thirds of the training steps (compared to training with statistics pooling) are required for the correlation-based model to reach the plateau on the evaluation EER. As expected, the introduction of the channel dropout increases the number of training steps required to reach convergence. Overall, the training time required for models with the statistics pooling and correlation pooling with dropout are comparable.



\begin{figure}[t]
  \centering
  \includegraphics[width=\linewidth]{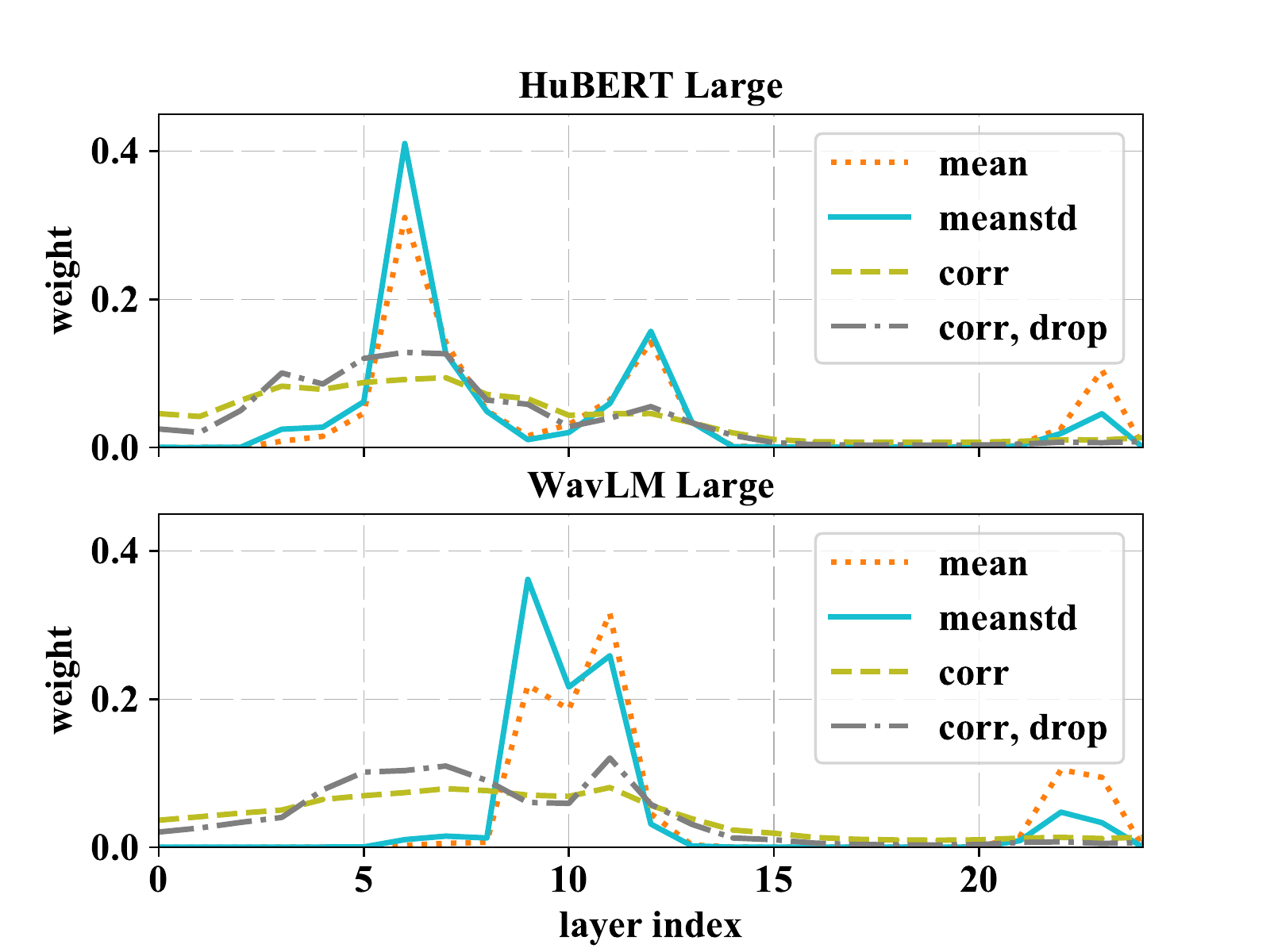}
  \vspace{-8mm}
  \caption{Learned weights per HuBERT-Large and WavLM-Large layers for SID.}
  \label{fig:sidhubertwavlmweights}
\end{figure}

\begin{figure}[ht]
  \centering
  \includegraphics[width=\linewidth]{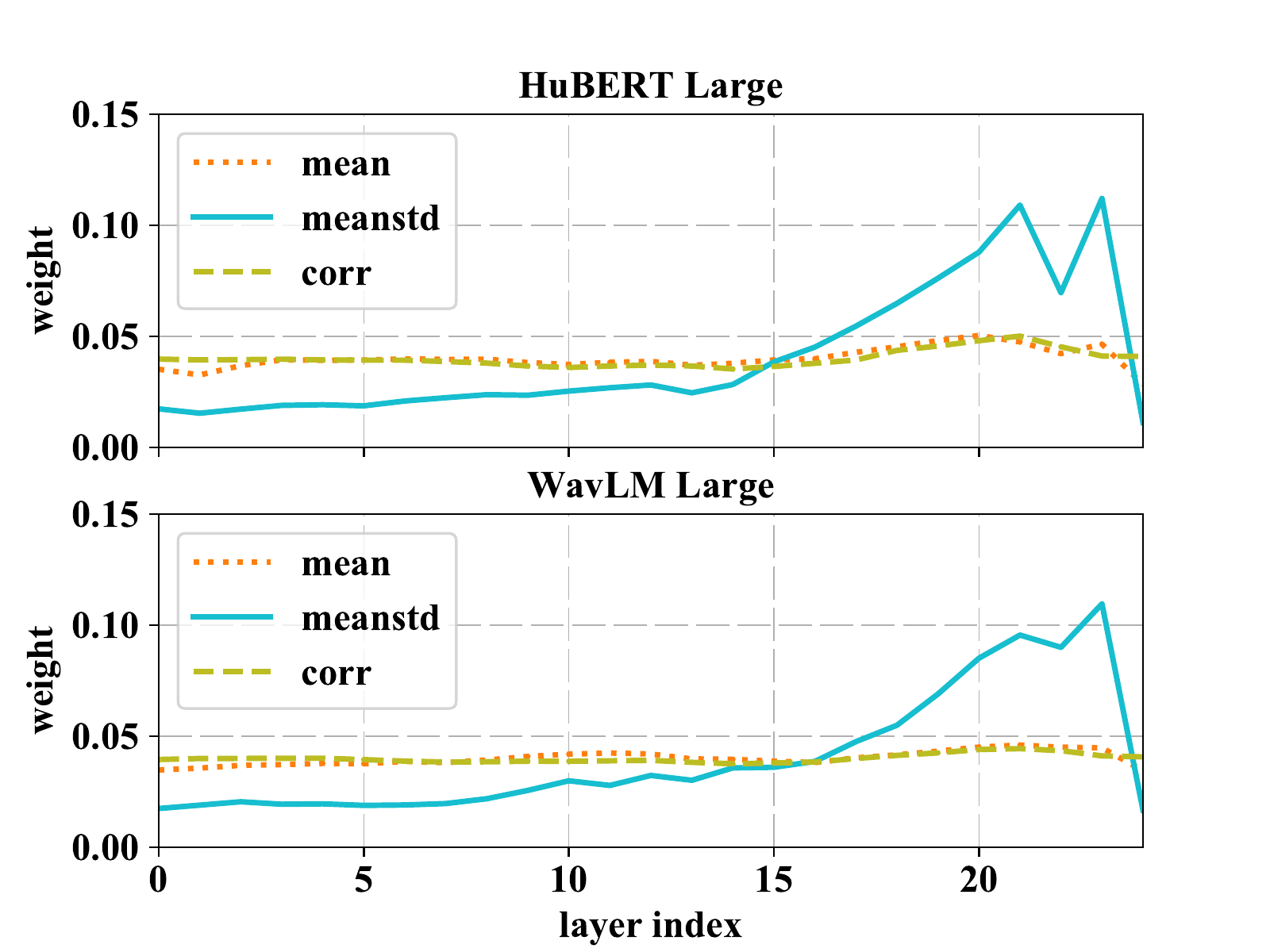}
  \vspace{-8mm}
  \caption{Learned weights per HuBERT-Large and WavLM-Large layers for ER.}
  \label{fig:erhubertwavlmweights}
  \vspace{-5mm}
\end{figure}

\vspace{-1em}
\subsection{Emotion recognition (ER)}
\label{ssec:emodetect}

Conveying emotions is an intrinsic aspect of human communication. This makes speech emotion recognition (ER) an important attribute for effective human-computer interaction and for this reason ER has attracted increasing attention. Emotional information resides beyond segmental productions and in longer time scales that involve the construction of fine-grained spectral cues across an utterance. As such, the performance of an emotion recognition system greatly depends on the representational capacity of engineered features to capture such information from audio. This has been typically done using a wide range of feature functionals such as eGeMAPS but also using MFCCs and filterbanks \cite{venkataramanan2019emotion, pepino2021emotion}.

\vspace{-3mm}
\subsubsection{ER Dataset}

\label{ssec:emodatasets}
The Interactive Emotional Dyadic Motion Capture (IEMOCAP) database consists of multi-modal recordings (speech and video) by 10 actors carried out in dyadic sessions in English and has a total duration of approximately 12 hours \cite{busso2008iemocap}. The actors performed both improvised hypothetical scenarios targeting the elicitation of specific emotions as well as selected emotional scripts.  The dataset is split in 5 dialogue sessions, each including a pair of male and female speakers. The emotions that the actors conveyed are happiness, anger, excitement, sadness, surprise, fear, frustration, and a neutral state. The dialogues were manually segmented at the speaker turn level and emotional labels were assigned on the basis of agreements from subjective emotional evaluations from human annotators. Similar to other studies, we follow the conventional evaluation protocol and relabel excitement samples as happiness and use 4 balanced emotional classes, namely: anger, happiness, sadness, and neutral \cite{yang2021superb, pepino2021emotion, fayek2017evaluating}. All other classes are discarded. Similar to \cite{yang2021superb}, we use Session1 for testing and the remaining data for training and validation. This approach leaves approximately 19\% of the data for testing.

\vspace{-3mm}
\subsubsection{ER Results}
\label{ssec:emoresults}

An overview of the results is presented in Table \ref{tab:emodetectionresults}. As we observe, correlation pooling yields better accuracy compared to mean pooling, which is the pooling method in SUPERB. When compared to mean-std pooling, the performance is similar (better than mean-std with WavLM and slightly lower with HuBERT; better than mean-std in both cases when correlation is used with dropouts). Fusion is not harmful, although the gains seem to be negligible. Finally, the learned weights for each pooling method are shown in Fig. \ref{fig:erhubertwavlmweights}.

\vspace{-3mm}
\subsection{Complementarity of pooling methods}
\label{ssec:complementarity}

In the SV task, given the TDNN-based downstream model, Fig. \ref{fig:svhubertwavlmweights} suggests that learned layer weights are, to some extent, agnostic to the type of pooling. Therefore, the inputs to the downstream speaker extractors employing different information pooling are similar. Since the SV results improve when fusing systems with mean and correlation pooling, we argue that the utilization of various pooling methods leads to models exhibiting complementarity of predictions.

On the other hand, layer weights learned when using mean- and correlation-based pooling in the context of SID are significantly different (see Fig. \ref{fig:sidhubertwavlmweights}). Hence, the improvements provided by fusion may be attributed to the difference between pooling methods as well as layer weights. To isolate the effect of weights, we performed the following experiment. The weights of the
layer-wise pooling $\{\gamma_l\}_{l=0}^L$ which are estimated using the ``SID model with correlation pooling with dropout'' are
copied and used as frozen weights when training the ``SID model using mean pooling''. The results are reported in Table \ref{tab:sidcomplementarity}, together with some fusion results with the new models. We observe the following: (a)~Correlation pooling used during optimization can steer the training towards finding more informative representations through layer weights. Using such weights leads to better results (especially with WavLM Large) when subsequently training with incorporated mean pooling. (b)~Even though mean-pooling-based models improved, they still do not outperform those trained with the correlation pooling from scratch. (c)~We confirmed the complementarity of the pooling approaches as the fusion of different systems --- with enforced identical inputs --- yields significant improvements (last row of Table \ref{tab:sidcomplementarity}).

\begin{table}[tb]
\centering
\caption{Unweighted accuracy (\%) for the emotion  recognition task in IEMOCAP using HuBERT and WavLM self-supervised representations.}
\label{tab:emodetectionresults}
\begin{tabular}{|p{3cm}||P{1.5cm}|P{1.5cm}|}
 \hline
 Pooling method & HuBERT Large & WavLM Large \\
 \hline
 mean   & 60.46 & 66.73 \\
 meanstd &   68.57  & 69.86 \\
 correlation & 68.21 & 71.43 \\
 corr. w/ dropout & 69.95 & 71.43\\
 mean+correlation & 68.94 & 71.43\\
 meanstd+correlation & 69.22 & 70.41\\
 \hline
\end{tabular}
\vspace{-6mm}
\end{table}

\begin{table}[tb]
\centering
\caption{Comparison of SID systems accuracy (\%) employing mean pooling (\emph{mean}); mean pooling and layer weights discovered by correlation pooling with dropout (\emph{mean}$^*$).}
\label{tab:sidcomplementarity}
\begin{tabular}{|p{3.5cm}||P{1.5cm}|P{1.5cm}|}
 \hline
 Approach & HuBERT Large & WavLM Large \\
 \hline
 mean   & 89.3 & 93.0 \\
 mean$^*$ & 91.7 & 97.2 \\
 corr. w/ drop. &  95.3 & 97.7 \\
 \hline
 mean + corr. w/ drop. & 96.2 & 98.3 \\
 mean$^*$ + corr. w/ drop. & 96.4 & 98.7 \\
 \hline
\end{tabular}
\vspace{-5mm}
\end{table}

\vspace{-1mm}
\section{CONCLUSIONS AND FUTURE DIRECTIONS}
\vspace{-2mm}
\label{sec:print}
In this paper, we developed and evaluated a pooling method that uses correlations between channels to extract information relevant to the task in hand. We followed the recently introduced SUPERB benchmark, where the goal is to examine the effectiveness of different self-supervised models in several tasks. Among them, we chose SID, SV, and ER and showed that (a) in speaker-related tasks, the proposed pooling is clearly superior to mean-std (statistics) pooling, and (b) in emotion recognition, the proposed method is superior to mean pooling (used in SUPERB) and comparable to mean-std pooling. Finally, we showed that fusion between different pooling methods can yield further improvements. 

These findings shed some light on the way self-supervised architectures capture certain utterance-level characteristics of speech, such as speaker, emotion, and potentially others. Note that the correlation is complementary to the mean and std (e.g. a Multivariate Normal Distribution can be parametrized using these 3 statistical quantities). Therefore, the information captured by it cannot be captured by the other two quantities, at least when the pretrained model is frozen. By exploring the complementarity, we achieved the best results in SID with WavLM Large and mean pooling, 97.2\%, by using pretrained and fixed layer weights estimated with correlation pooling (the current best result in the leaderboard is 95.5\%).

As future directions, we propose to combine correlation pooling with attention, evaluate it on other tasks (e.g. channel and spoken language recognition), employ the intrinsic geometry of correlation matrices instead of the Euclidean (e.g. using methods proposed in \cite{thanwerdas2021geodesics}), and examine its performance after fine-tuning the backbone network. 

\vspace{-2mm}
\section{Acknowledgements}
\vspace{-2mm}
The work was supported by Czech Ministry of Interior project No. VJ01010108 "ROZKAZ", Czech National Science Foundation (GACR) project NEUREM3 No. 19-26934X, Czech Ministry of Education, Youth and Sports project no. LTAIN19087 "Multi-linguality in speech technologies", and Horizon 2020 Marie Sklodowska-Curie grant ESPERANTO, No. 101007666. Computing on IT4I supercomputer was supported by the Czech Ministry of Education, Youth and Sports from the Large Infrastructures for Research, Experimental Development and Innovations project "e-Infrastructure CZ – LM2018140". Sofoklis Kakouros is supported by the Academy of Finland through project no. 340125.
\bibliographystyle{IEEEbib}
\bibliography{refs}

\begin{thebibliography}{10}

\bibitem{hsu2021hubert}
Wei-Ning Hsu, Benjamin Bolte, Yao-Hung~Hubert Tsai, Kushal Lakhotia, Ruslan
  Salakhutdinov, and Abdelrahman Mohamed,
\newblock ``{HuBERT: Self-supervised speech representation learning by masked
  prediction of hidden units},''
\newblock {\em IEEE/ACM Transactions on Audio, Speech, and Language
  Processing}, vol. 29, pp. 3451--3460, 2021.

\bibitem{fan2020exploring}
Zhiyun Fan, Meng Li, Shiyu Zhou, and Bo~Xu,
\newblock ``Exploring wav2vec 2.0 on speaker verification and language
  identification,''
\newblock {\em arXiv preprint arXiv:2012.06185}, 2020.

\bibitem{chen2021wavlm}
Sanyuan Chen, Chengyi Wang, Zhengyang Chen, Yu~Wu, Shujie Liu, Zhuo Chen, Jinyu
  Li, Naoyuki Kanda, Takuya Yoshioka, Xiong Xiao, et~al.,
\newblock ``{WavLM: Large-scale self-supervised pre-training for full stack
  speech processing},''
\newblock {\em arXiv preprint arXiv:2110.13900}, 2021.

\bibitem{yang2021superb}
Shu-wen Yang, Po-Han Chi, Yung-Sung Chuang, Cheng-I~Jeff Lai, Kushal Lakhotia,
  Yist~Y Lin, Andy~T Liu, Jiatong Shi, Xuankai Chang, Guan-Ting Lin, et~al.,
\newblock ``{SUPERB: Speech processing universal performance benchmark},''
\newblock in {\em Proceedings of Interspeech}, 2021.

\bibitem{snyder2018xvec}
David Snyder, Daniel Garcia-Romero, Gregory Sell, Daniel Povey, and Sanjeev
  Khudanpur,
\newblock ``{X-Vectors: Robust DNN Embeddings for Speaker Recognition},''
\newblock in {\em 2018 IEEE International Conference on Acoustics, Speech and
  Signal Processing (ICASSP)}, 2018, pp. 5329--5333.

\bibitem{wang2021pooling}
Shuai Wang, Yexin Yang, Yanmin Qian, and Kai Yu,
\newblock ``Revisiting the statistics pooling layer in deep speaker embedding
  learning,''
\newblock in {\em 2021 12th International Symposium on Chinese Spoken Language
  Processing (ISCSLP)}, 2021, pp. 1--5.

\bibitem{stafylakis2021speaker}
Themos Stafylakis, Johan Rohdin, and Lukas Burget,
\newblock ``Speaker embeddings by modeling channel-wise correlations,''
\newblock in {\em Interspeech}, 2021.

\bibitem{gatys2016image}
Leon~A Gatys, Alexander~S Ecker, and Matthias Bethge,
\newblock ``Image style transfer using convolutional neural networks,''
\newblock in {\em Proceedings of the IEEE conference on computer vision and
  pattern recognition}, 2016, pp. 2414--2423.

\bibitem{chorowski2018using}
Jan Chorowski, Ron~J Weiss, Rif~A Saurous, and Samy Bengio,
\newblock ``On using backpropagation for speech texture generation and voice
  conversion,''
\newblock in {\em 2018 IEEE International Conference on Acoustics, Speech and
  Signal Processing (ICASSP)}. IEEE, 2018, pp. 2256--2260.

\bibitem{vaswani2017transformer}
Ashish Vaswani, Noam Shazeer, Niki Parmar, Jakob Uszkoreit, Llion Jones,
  Aidan~N Gomez, \L~ukasz Kaiser, and Illia Polosukhin,
\newblock ``{Attention is All you Need},''
\newblock in {\em Advances in Neural Information Processing Systems}, I.~Guyon,
  U.~Von Luxburg, S.~Bengio, H.~Wallach, R.~Fergus, S.~Vishwanathan, and
  R.~Garnett, Eds. 2017, vol.~30, Curran Associates, Inc.

\bibitem{peters-etal-2018-deep}
Matthew~E. Peters, Mark Neumann, Mohit Iyyer, Matt Gardner, Christopher Clark,
  Kenton Lee, and Luke Zettlemoyer,
\newblock ``{Deep Contextualized Word Representations},''
\newblock in {\em Proceedings of the 2018 Conference of the North {A}merican
  Chapter of the Association for Computational Linguistics: Human Language
  Technologies, Volume 1 (Long Papers)}, New Orleans, Louisiana, June 2018, pp.
  2227--2237, Association for Computational Linguistics.

\bibitem{novoselov2022robust}
Sergey Novoselov, Galina Lavrentyeva, Anastasia Avdeeva, Vladimir Volokhov, and
  Aleksei Gusev,
\newblock ``{Robust Speaker Recognition with Transformers Using wav2vec 2.0},''
\newblock {\em arXiv preprint arXiv:2203.15095}, 2022.

\bibitem{van2022comparison}
Benjamin van Niekerk, Marc-Andr{\'e} Carbonneau, Julian Za{\"\i}di, Matthew
  Baas, Hugo Seut{\'e}, and Herman Kamper,
\newblock ``A comparison of discrete and soft speech units for improved voice
  conversion,''
\newblock in {\em ICASSP 2022-2022 IEEE International Conference on Acoustics,
  Speech and Signal Processing (ICASSP)}. IEEE, 2022, pp. 6562--6566.

\bibitem{nagrani2017voxceleb}
Arsha Nagrani, Joon~Son Chung, and Andrew Zisserman,
\newblock ``{VoxCeleb: A Large-Scale Speaker Identification Dataset},''
\newblock {\em Proc. Interspeech 2017}, pp. 2616--2620, 2017.

\bibitem{sadjadi2021nist}
Omid Sadjadi,
\newblock ``{NIST SRE CTS Superset: A large-scale dataset for telephony speaker
  recognition},''
\newblock {\em arXiv preprint arXiv:2108.07118}, 2021.

\bibitem{greenberg2020nistevals}
C.~Greenberg, S.~O. Sadjadi, L.~Mason, and D.~Olson,
\newblock ``{Two Decades of Speaker Recognition Evaluation at the National
  Institute of Standards and Technology},''
\newblock {\em Computer Speech and Language}, 2020.

\bibitem{lavrentyeva2022stc}
Galina Lavrentyeva, Sergey Novoselov, Vladimir Volokhov, Anastasia Avdeeva,
  Aleksei Gusev, Alisa Vinogradova, Igor Korsunov, Alexander Kozlov, Timur
  Pekhovsky, Andrey Shulipa, et~al.,
\newblock ``{STC Speaker Recognition System for the NIST SRE 2021},''
\newblock in {\em Proc. The Speaker and Language Recognition Workshop}, 2022.

\bibitem{vaessen2022fine}
Nik Vaessen and David~A Van~Leeuwen,
\newblock ``Fine-tuning wav2vec2 for speaker recognition,''
\newblock in {\em ICASSP 2022-2022 IEEE International Conference on Acoustics,
  Speech and Signal Processing (ICASSP)}. IEEE, 2022, pp. 7967--7971.

\bibitem{chung2018voxceleb2}
Joon~Son Chung, Arsha Nagrani, and Andrew Zisserman,
\newblock ``{VoxCeleb2: Deep Speaker Recognition},''
\newblock {\em Proc. Interspeech 2018}, pp. 1086--1090, 2018.

\bibitem{wang2018am}
Feng Wang, Jian Cheng, Weiyang Liu, and Haijun Liu,
\newblock ``{Additive Margin Softmax for Face Verification},''
\newblock {\em IEEE Signal Processing Letters}, vol. 25, no. 7, pp. 926--930,
  2018.

\bibitem{venkataramanan2019emotion}
Kannan Venkataramanan and Haresh~Rengaraj Rajamohan,
\newblock ``Emotion recognition from speech,''
\newblock {\em arXiv preprint arXiv:1912.10458}, 2019.

\bibitem{pepino2021emotion}
Leonardo Pepino, Pablo Riera, and Luciana Ferrer,
\newblock ``Emotion recognition from speech using wav2vec 2.0 embeddings,''
\newblock {\em arXiv preprint arXiv:2104.03502}, 2021.

\bibitem{busso2008iemocap}
Carlos Busso, Murtaza Bulut, Chi-Chun Lee, Abe Kazemzadeh, Emily Mower, Samuel
  Kim, Jeannette~N Chang, Sungbok Lee, and Shrikanth~S Narayanan,
\newblock ``{IEMOCAP: Interactive emotional dyadic motion capture database},''
\newblock {\em Language resources and evaluation}, vol. 42, no. 4, pp.
  335--359, 2008.

\bibitem{fayek2017evaluating}
Haytham~M Fayek, Margaret Lech, and Lawrence Cavedon,
\newblock ``Evaluating deep learning architectures for speech emotion
  recognition,''
\newblock {\em Neural Networks}, vol. 92, pp. 60--68, 2017.

\bibitem{thanwerdas2021geodesics}
Yann Thanwerdas and Xavier Pennec,
\newblock ``{Geodesics and Curvature of the Quotient-Affine Metrics on
  Full-Rank Correlation Matrices},''
\newblock in {\em International Conference on Geometric Science of
  Information}. Springer, 2021, pp. 93--102.

\end{thebibliography}

\end{document}